\pdfoutput=1
\documentclass[11pt]{article}
\usepackage[utf8]{inputenc}
\usepackage[T1]{fontenc}
\usepackage{fixltx2e}
\usepackage{graphicx}
\usepackage{longtable}
\usepackage{float}
\usepackage{wrapfig}
\usepackage{rotating}
\usepackage[normalem]{ulem}
\usepackage{amsmath}
\usepackage{textcomp}
\usepackage{marvosym}
\usepackage{wasysym}
\usepackage{amssymb}
\usepackage{hyperref}
\usepackage{microtype}
\graphicspath{{figures/}}
\usepackage{wrapfig}
\restylefloat{figure}
\usepackage[margin=1in]{geometry}
\usepackage[backref=true,backend=biber,sorting=none,citestyle=numeric-comp]{biblatex}
\author{B.L. DeCost, J.R. Hattrick-Simpers, Z. Trautt, A.G. Kusne, E. Campo, M.L. Green}
\date{}
\title{Scientific AI in Materials Science: a Path to a Sustainable and Scalable Paradigm}
\hypersetup{
  pdfkeywords={},
  pdfsubject={},
  pdfcreator={Emacs 25.3.1 (Org mode 8.2.10)}}
\begin{document}

\maketitle
\begin{abstract}
Recently there has been an ever-increasing trend in the use of machine
learning (ML) and artificial intelligence (AI) methods by the materials
science, condensed matter physics, and chemistry communities. This
perspective article identifies key scientific, technical, and social
opportunities that the materials community must prioritize to
consistently develop and leverage Scientific AI (SciAI) to provide a
credible path towards the advancement of current materials-limited
technologies. Here we highlight the intersections of these opportunities with a series
of proposed paths forward. The opportunities are roughly sorted from
scientific/technical (\emph{e.g.} development of robust, physically
meaningful multiscale material representations) to social (\emph{e.g.}
promoting an AI-ready workforce). The proposed paths forward range from
developing new infrastructure and capabilities to deploying them in
industry and academia. We provide a brief introduction to AI in
materials science and engineering, followed by detailed discussions of
each of the opportunities and paths forward.
\end{abstract}

\begin{wrapfigure}{R}{0.4\textwidth}
  \includegraphics[width=0.4\textwidth]{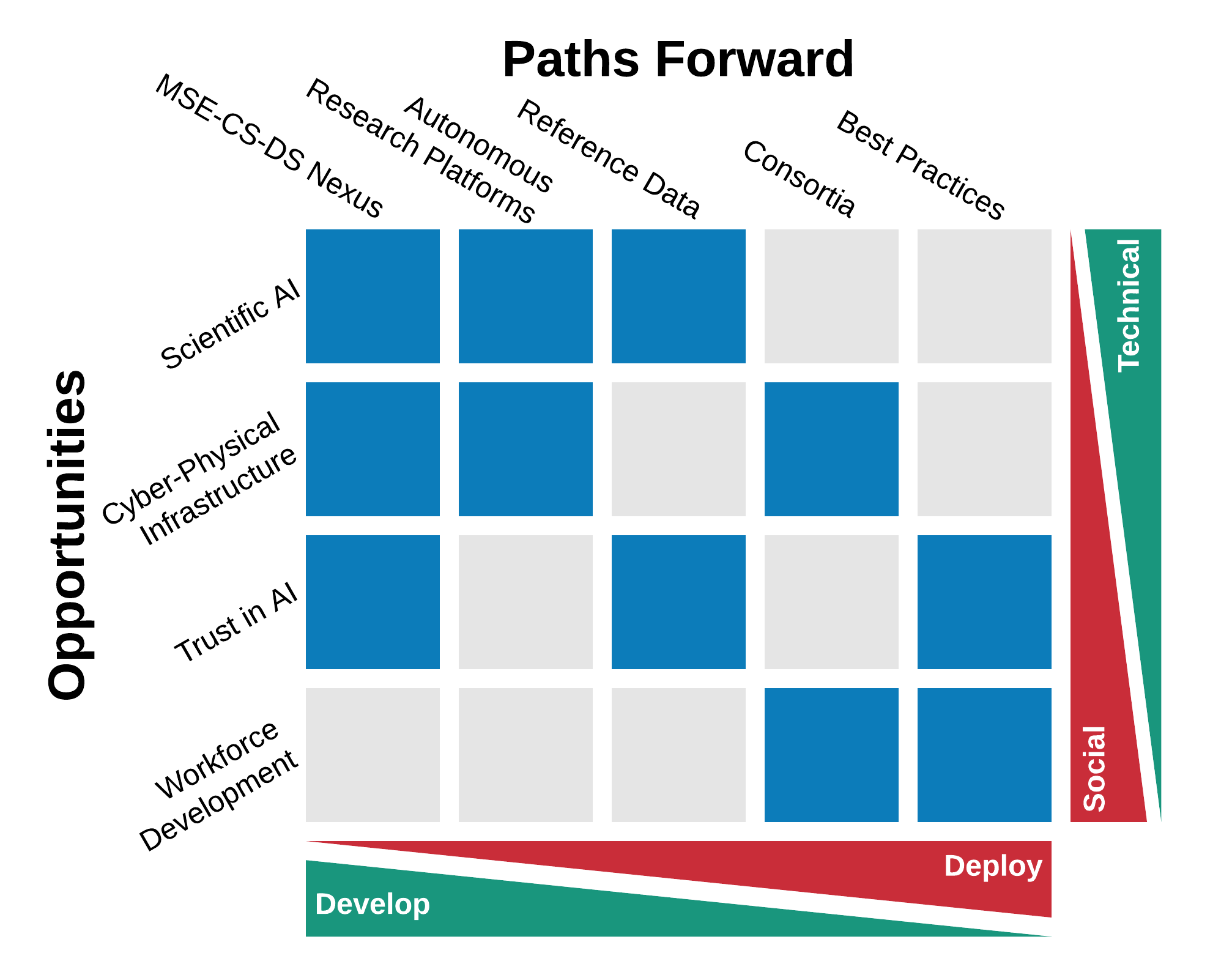}
  \caption{Opportunities and proposed paths forward that will enable
broader and more effective use of AI in materials science and
engineering. The blue intersections represent areas of particular
importance to be discussed in this paper.}
\end{wrapfigure}

\section{A Brief perspective on AI in materials science}
\label{sec-1}

Recent reports, reviews, symposia, and workshops have heralded machine
learning (ML) and artificial intelligence (AI) methods as the next
scientific paradigm in materials discovery and optimization \cite{agrawal16_persp,kalidindidegraef,dimiduk2018}.
Applications to materials science have exploded, spanning data analysis, knowledge
extraction, and experiment selection \cite{holdren2014,MAP,DOE2017,DOE2019,agrawal16_persp}.
The numerous reasons for this trend are
related to the omnipresence of ML systems in our everyday lives, the
free availability software, and the demonstrated successes in materials
discovery and on-the-fly data acquisition inspired by the Materials
Genome Initiative (MGI) \cite{aziza2018,lookman2016,ren2018,agrawal16_persp}.
However, despite their recent prominence, these
techniques have been applied in a variety of materials science fields
since the early 1960's \cite{Hussey1963,DeWilde1987,Buratijr.1983,Teti1994,Bhadeshia_1999}.

Some recent examples of the successful implementation of ML to materials
science were demonstrated by the high-throughput experimental (HTE, also
known as “combinatorial”) community. Here a critical data analytics
bottleneck resulted from having hundreds to thousands of high-quality
measurements correlated in composition, processing and microstructure \cite{long2007,long2009,kusne2014,suram2016}.
There have been several international
efforts to standardize data formats and create data analysis and
interpretation tools for large scale data sets \cite{koinuma2002,lippmaa2005,chikyow2006}. The rise of the
HTE community resulted in the creation of new and creative modes of
measuring properties and visualizing/interpreting data, which will also
accelerate materials innovation in emerging AI-enabled autonomous
methods \cite{nikolaev16_auton_mater_resear,sanchez-lengeling18_inver_molec_desig_using_machin_learn,dunn2019rocketsled,talapatra2018}.

Although the application of AI is now increasingly commonplace in the
materials community, we are approaching the peak of excitement and
inflated expectations. A downturn of disillusionment is inevitable, but
we believe that by pursuing the following opportunities the community
will more rapidly reach a steady state of AI usefulness. Achieving this
goal will require 1) methodological development in scientific AI, 2)
significant investment in cyber-physical infrastructure, 3) commitment
to measures improving trust in AI systems, and 4) workforce development.

\section{Opportunities in Scientific AI}
\label{sec-2}

The transition from expectations to practice for AI will require
development of robust Scientific AI systems that can go beyond
generating leads, i.e., nudges in the right direction, to providing rich
functionality that enables scientific discovery. Two opportunities to
close this gap are:

\begin{enumerate}
\item developing Scientific AI systems that combine ML techniques with physical mechanisms
\item innovative applications of AI systems to directly derive scientific insight
\end{enumerate}

A robust community of interdisciplinary materials science and
engineering (MSE) and ML researchers is needed to enable the algorithmic
development to support these two goals. Distributed automated laboratory
systems will facilitate this development by equalizing access to cutting
edge experimental materials science, providing a substrate for
high-impact interdisciplinary collaboration. Materials have always been
technology enablers, and currently there are many key technology areas
that await materials discovery and processing solutions (reference).
Addressing these opportunities will drive and propel the required
developments.

\subsection{Getting physical mechanisms into ML models}
\label{sec-2-1}

The brute force strategy of collecting massive annotated datasets, such
as those that enabled the current wave of advances in image recognition,
natural language understanding, and neural translation, is untenable due
to the relative scarcity of many types of materials data, and the high
cost of obtaining materials data. One of the greatest opportunities to
expand application of AI methods is in addressing underdeveloped
material and processing representations. Simple models fail to capture
the complexity of hierarchical materials structures (i.e., they
underfit), while high capacity models often yield pathological or
trivial results for small and medium-sized datasets (i.e., they
overfit). This results from the classic bias-variance tradeoff \cite{kohavi1996bias}.

An alternative strategy for dealing with complex modeling problems is to
deliberately introduce the \textbf{right} kind of bias into high capacity
models by designing input representations and model forms to reflect
known invariances, equivariances, and symmetries in the domain \cite{anselmi2019}. In
the context of scientific AI, this means incorporation of \textbf{mechanistic
biases} to create interpretable models and learning algorithms,in the
form of physical heuristics, theories, and laws. For example, the
Physically Inspired Neural Network interatomic potential \cite{pun2019} uses a neural network to adaptively parametrize a
classical interatomic potential form instead of directly modeling forces
and energies.

A key opportunity is to systematically integrate the vast implicit and
explicit materials knowledge in the published literature on a per-task
basis through model form specification and learning algorithm design
choices. This principle is applied in a limited way in the materials AI
community, but much research is needed to more fully incorporate
physical intuition before ML models can extrapolate to new regions of
material space. This is an area where the development of enterprise
knowledge-graphs and ontologies that capture subject matter expertise
will help to provide more actionable features.

In addition to incorporating mechanistic biases at the level of
individual modeling tasks, scientific AI systems for materials
development will require the development of hybrid machine learning
systems that bridge time and length scales as well as experimental and
computational paradigms. Outside of the interatomic potentials
community, there are few demonstrations of representing material
structure representations tailored for dynamic processes. It is
difficult to encode certain types of metadata (environment, processing
paths, heat transfer characteristics that depend on geometry, etc.) that
are known to influence material properties. Vector-valued and
time-varying material processing attributes (such as loading and
annealing schedules) are often reduced to categorical and tabular
representations. Importantly, much effort is required to address
technologically important materials systems, where the complexity of
material processing far exceeds that of laboratory-scale studies.

\subsection{Deriving scientific insight from AI models}
\label{sec-2-2}

In many ways, current applications of AI in materials science focus more
on solving engineering and design problems than on directly deriving
fundamental scientific insight from data. Current materials science AI
applications predominantly focus on lead generation and black-box
optimization. To realize the full potential of AI to help us more
efficiently and effectively practice scientific inquiry, the materials
community must develop AI systems that can represent, evaluate, and
perform inference about physical mechanisms underlying observational
data.

In the short term, creative application of existing ML methods is
enabling new avenues to accelerate scientific discovery. Active
learning, for example, might be applied to identify a set of optimal
experiments to disambiguate a list of potential physical theories, as is
being explored in the social sciences \cite{ouyang16}. Similarly,
algorithmically driven experimentation could be used to search for
counterexamples to heuristic models or physical theories, potentially
providing materials scientists with valuable insights into why these
heuristics and theories break down \cite{jia2019anthropogenic}. Furthermore, much
of the existing materials knowledge base is in the form of implicit
institutional knowledge and expert intuition. Thus, development of
“human-in-the-loop” methodologies leveraging real-time model
visualization, introspection, and feedback must not be overlooked.

An important next step in scientific AI is the development of new AI
methods tailored for scientific discovery. This includes methods that
can infer physical relationships, mechanisms, and principles from data,
potentially drawing from the fields of causal discovery \cite{heckerman1999bayesian} and
probabilistic programming \cite{vajda2014probabilistic}. At the “Strong AI” extreme of this line
of inquiry, hypothetical AI systems will be expected to formulate and
test scientific theories to credibly identify new scientific paradigms.
Even if such systems can be constructed, they will still face the
\emph{Pauling Problem} \cite{shechtman2013quasi}, where physical bias overwhelms new evidence of
world-view-breaking phenomena such as superconductivity, 2D materials,
or quasicrystals.

\subsection{Paths forward}
\label{sec-2-3}

\textbf{Cross-disciplinary Collaboration}
\begin{itemize}
\item Generate funding opportunities targeted towards funding cross-disciplinary research at the cutting edge of MSE, ML, AI, and Robotics to promote communication skills to identify and frame mutually interesting research.
\item Collaborate to develop multiscale materials and knowledge representations and  generative modeling techniques
\item Create career opportunities at the research associate and technician levels in applied ML and Software Engineering.
\item Explore probabilistic programming methods to meld physical and phenomenological modeling with machine learning.
\item Develop objective methods for identification and evaluation of the most informative or unusual datum in any given scientific dataset.
\end{itemize}

\textbf{Autonomous research platforms:}
\begin{itemize}
\item Develop open autonomous research platforms to provide a substrate for developing and deploying materials AI methods on large-scale materials design problems.
\item Provide opportunities for the broad materials and AI communities to have access to these platforms, lowering the barrier to entry to materials discovery and design.
\end{itemize}

\textbf{Reference data:}
\begin{itemize}
\item Develop challenge problems to focus innovation and collaboration on  difficult scientific discovery problems, i.e. the materials discovery and design analog to Large Scale Visual Recognition Challenge \cite{Russakovsky_2015}.
\item Compile materials datasets with annotated physical rules and heuristics.
\end{itemize}

\section{Opportunities in Cyber-Physical infrastructure}
\label{sec-3}

Realization of scientific AI's potential in materials science and
engineering will require advanced cyber-physical infrastructure. We have
identified four major opportunities to facilitate this development:

\begin{enumerate}
\item Improved standards and coordination in materials data infrastructure
\item Development of open and interoperable API-enabled experimental tools
\item Development of scalable on-demand synthesis/characterization capabilities
\item Democratization of research platforms
\end{enumerate}

An improved materials data infrastructure will enable data stewardship
throughout the research data lifecycle, which will greatly improve the
accessibility of data and metadata to both AI systems and human
researchers. Fully automate-able synthesis and characterization tools
that execute standardized experimental protocols will improve
reproducibility while seamlessly capturing provenance. This will
decrease the cost of generating creating new data and knowledge, and
will support real-time distributed and autonomous experimentation.
Development of new impedance matched, on-demand synthesis and
characterization techniques will be critical to expand the applicability
of this approach. The fundamental question is how do we rethink the
“synthesize-then-characterize” framework when actionable knowledge can
be generated at a rate faster than it takes to transfer the specimens?
Finally, we must develop organizational frameworks to democratize access
to these new experimental, computational, and data resources, something,
comparable to the user facility paradigm at high performance computing
centers. Ultimately, this framework would enable scientists and
engineers to focus more of their time on conceiving, planning, and
executing scientific studies.

\subsection{Standards and coordination in materials data infrastructure}
\label{sec-3-1}

Over the past decade, several reports have identified materials data
infrastructure as critical gaps limiting innovation in materials
research \cite{tinkle2013sharing,ward2015materials,jain2016mgi}. These reports consistently highlight the need for
long-term support of shared data services, improved coordination among
government agencies, publication of all research data (novel as well as
null) with robust metadata, and improved development of community
standards for these data and metadata. Findable, Accessible,
Interoperable, and Reusable (FAIR) data principles \cite{wilkinson2016fair}  can guide the
materials science and engineering community
in developing infrastructure suited to collaborative and adaptive
research. However, the complexity of materials science and engineering
data poses unique challenges to the adoption of FAIR principles.
International groups, such as the Research Data Alliance, are fully
embracing FAIR Data Principles and are extending them beyond data and
metadata, to data types, instruments, and physical samples. Currently,
the materials science and engineering community does not have robust
frameworks for assigning persistent identifiers to data types,
instruments, physical samples, and data and metadata within a larger
dataset. Furthermore, once persistent identifiers are assigned on
smaller units within a larger dataset, the community will face
challenges in effectively and uniformly citing data.

\subsection{Open and interoperable API-enabled experimental tools}
\label{sec-3-2}

Critical bottlenecks for adaptive science and autonomous control of
experimental systems are (i) a widespread absence of application
programming interfaces (API) to interact with laboratory equipment, (ii)
lack of a unified language for experimental workflow protocols, and
(iii) lack of standardized and open data formats to facilitate
accessibility and interoperability. Currently, downstream researchers
are developing \emph{ad hoc} hardware interfaces, duplicating effort and
often incurring substantial technical debt. Materials synthesis and
characterization workflows are typically manifested in custom software
rather than in composable and machine-actionable data representations.
Finally, experimental equipment is supported by a diverse collection of
vendor-specific interfaces and formats, which may not be
well-documented, and may be difficult to use independently from
vendor-developed software frameworks. This presents an unnecessary
impediment to innovation in real-time data analysis and adaptive
experimental planning and control. There is a significant need to
facilitate industry-lead development of standards for open and
machine-actionable instrument APIs, executable protocols for
experimental workflows, and file formats.

\subsection{Scalable on-demand synthesis/characterization capabilities}
\label{sec-3-3}

Current materials synthesis and characterization tools are not designed
for low latency and high agility between experiments, leading to a
significant time-constant mismatch with the algorithmic hypothesis
generation enabling autonomous experimentation. While high-throughput
synthesis techniques enabled revolutionary improvements in the rapid
exploration of process-structure-property relationships [???], library
generation now presents a major bottleneck due to its high latency and
the intensity of human labor involved. Therefore, low latency, automated
synthesis platforms, integrated with multimodal characterization tools,
should be developed. AI also presents unprecedented opportunities for
novel adaptive experiments enabled by \emph{in situ} automated perception and
data analysis, e.g., through real-time identification, tracking, and
subsequent fine-grained analysis of features of interest \cite{burnett2019completing}. For
low-latency decision-making, it may be necessary to leverage edge
computing \cite{shi2016edge}, e.g. running a deep learning model directly on detector
output.

\subsection{Resource Democratization}
\label{sec-3-4}

Large materials research user facilities (e.g. Advanced Photon Source,
NERSC) have demonstrated a model for decoupling the construction and
operation of experimental tools and computing infrastructure from the
use of those tools by scientific subject matter experts. Similarly, the
adaptive synthesis and integrated multimodal characterization platforms
described above will require significant capital investment to invent,
develop, and build. Therefore, the materials community, and the greater
community at large, is presented with an opportunity to develop an
organizational and technological framework to facilitate collaboration
between theoretical and experimental research groups, and to lower the
barrier for cross-material-system, cross-synthesis-method, and
cross-modality studies. This framework would also provide increased
access to cutting edge experimental materials capabilities to new user
communities from underrepresented groups and smaller institutions. In
addition to the cyber-physical infrastructure challenges described
above, experimental synthesis and characterization methods are very
specific to a given class of materials. There is unlikely to be even one
brick and mortar facility to allow researchers to study several
materials classes.

\subsection{Paths Forward}
\label{sec-3-5}

\textbf{Consortia}

\begin{itemize}
\item Develop community standards to enable FAIR data and equipment interoperability, while learning from successful examples, such as MTConnect
\item Design, deploy, operate, and provide democratized access to distributed autonomous laboratory platforms and broader cyber-physical infrastructure, as advocated in the high throughput experimental materials collaboratory (HTE-MC) concept \cite{HTEMC}.
\item Launch new funding initiatives to support creation of materials-focused AI Research Centers and Mission-Driven AI Laboratories as described in \cite{gil201920}.
\end{itemize}

\textbf{Autonomous materials science}
\begin{itemize}
\item Design for automation: Rethink the “high throughput” materials synthesis methodology portfolio in light of new capabilities in  real-time automated perception, modeling, decision making, and the need for real-time closed-loop feedback from multiple structure and  property probes.
\item Leverage automation: identify new opportunities to turn \emph{ex situ} analysis methods into AI-driven \emph{in situ} adaptive techniques.
\end{itemize}

\section{Opportunities in Trust}
\label{sec-4}

Promoting community-wide trust in Scientific-AI results is key to
reducing the impact of the downturn of disillusionment. We have
identified three important opportunities for improving confidence in
scientific AI as applied to materials:

\begin{enumerate}
\item Develop and enforce community wide standards for reporting uncertainty from archival data to final model predictions.
\item Create a scientific culture that values and promotes reproducibility, validation, and verification of published data
\item Work towards improving the interpretability of AI models and providing a solid foundation towards trust in their predictions
\end{enumerate}

Creating a robust interdisciplinary MSE-CS-DS nexus will create
opportunities for real-time methods for exploring materials
representations, permitting researchers to have confidence that the
final model reflects solid physical principles. Generation of reference
data sets and materials data challenge problems will allow benchmarking
of new models/algorithms using specially designed performance indicators
relevant to the specific AI task. Dissemination of best practices will
create a community of informed skeptics that request open code and
datasets, look for task-appropriate performance indicators, and are
alert to issues of dataset and modeling bias.

\subsection{Uncertainty: archival data to final model predictions}
\label{sec-4-1}

Current applications of AI in materials science largely ignore the
uncertainty of the raw data used to train models. Leveraging larger
scale datasets derived from the open literature and published materials
databases will require systematic evaluation of source and reporting
uncertainty. Reporting uncertainty is introduced by incomplete
collection, storage, and/or publication of relevant data, metadata,
experimental uncertainties, and potential spurious covariates. Even
where all metadata are recorded, fusing data from sources with different
intrinsic uncertainty levels presents an important modeling opportunity.
At any point where manual annotations from human experts (or
non-experts) enter the process, one must also account for annotation
uncertainty.

The outputs of ML models also have uncertainty related to the model
selection and fitting process. This kind of uncertainty must be
systematically propagated through a larger pipeline of interlinked
physical and ML models. Unbiased assessments of the full model
uncertainty from raw data through final predictions are needed to
determine with confidence whether it is reasonable to trust the
predictions of a machine learning pipeline. Furthermore, well-calibrated
uncertainty estimates are crucial to the performance of active learning
systems, which rely on quantification of model uncertainty to identify
experiments that are likely to be informative.

\subsection{Reproducibility, validation, and verification}
\label{sec-4-2}

Ensuring the reproducibility of scientific AI in materials research
depends critically on transparency in publication, attention to correct
methodology in evaluating results, and independent testing and
verification of model predictions. We must develop a strong culture of
scrutinizing modeling assumptions, checking for due diligence in
training procedures, and verifying that ML models are not being applied
outside their regime of applicability.

Recent development of open libraries (Matminer, TPOT), data repositories
and platforms (MP, AFLOW, OQMD, Citrination, MDF, NOMAD and AIIDA), and
paper repositories are significantly increasing the accessibility and
reproducibility of materials research. However, manuscripts often do not
fully document model hyperparameters, or the model selection and tuning
process used, and data and software are not commonly made available.
This can make it difficult to evaluate whether the results suffer from
overfitting or information leakage, and impossible for independent
verification or comparison with other works. Researchers using AI
methods should investigate and publish the failure modes of the models
they use, as this can promote improvement in, and trust of, AI methods.
For any given modeling task, choice of appropriate performance metrics
is of paramount importance \cite{riley2019three}. Metrics that account for dataset
bias are particularly important in the face of systemic publication bias
in favor of “positive” scientific results, community pursuit of “lead
material” derivatives, and in modeling phenomena governed by rare
materials features (such as fatigue crack initiation).

Finally, much of the materials AI literature describes proof-of-concept
work applied to a single material system, and experimental validation of
predictions is often deferred to followup studies. In contrast, many
computer science venues expect methods papers to demonstrate
generalizable results on multiple datasets and/or multiple tasks.
Addressing this problem will necessarily entail finding ways to lower
the barrier for groups to collaborate.

\subsection{Establishing Interpretability and Trust}
\label{sec-4-3}

Models and theories are fundamental to the scientific method, and
scientists expect to be able to rationalize predictions and discoveries
by explaining observations through an underlying phenomenon or
mechanism. Thus, the interpretability of scientific AI models is
necessary to establish sufficient trust in AI methods for widespread
scientific application. Interestingly, trust and interpretability
currently lack consensus definitions in computer science and
psychology \cite{schmidt2019quantifying}. The challenge of interpretability lies in balancing
faithful representation of the model's mechanisms and the ease of
intuitive understanding by a human \cite{herman2017promise}, while trust
corresponds to a user's willingness to accept or reject model
predictions relative to the baseline error rate of the model \cite{schmidt2019quantifying}.

The most common scenario in AI-driven materials science involves
completely opaque previously-generated models, for example in a
process-oriented environment \cite{holm2019defense}. Here the user does not have
access to the full descriptor set or material representation, may not
know the model form, and only has access to the final prediction. Thus
the distribution of user trust levels may have a large variance.
Informed trust in this scenario must be gained through meticulous
empirical validation procedures.

Feature importance ranking provides some level of insight into a model,
but does not go far enough to support claims of physical realism; this
interpretability tool is not typically robust in the face of correlated
or spurious inputs. Great opportunities exist to boost the
interpretability of AI models by providing output in the form of either
a human interpretable series of selection criteria (e.g. a simple
decision tree/process flow diagram) \cite{raccuglia2016}, a set of physically
meaningful equations, or a textual explanation. At this level of
interpretability expert opinions could be built transparently into the
framework through extensive interactions and the trust in the model
outputs will be increased.

\subsection{Paths Forward}
\label{sec-4-4}

\textbf{Cross-disciplinary Collaboration:}
\begin{itemize}
\item Develop and deploy real-time algorithms for exploring interpretable  material representations during research campaigns.
\item Design human-in-the-loop methodologies for quantifying interpretability and trust.
\end{itemize}

\textbf{Reference Data:}
\begin{itemize}
\item Develop and adopt common benchmark datasets and performance indicators for measuring and comparing methodological progress.
\item Create dedicated funding mechanisms for experimental validation of materials predictions
\end{itemize}

\textbf{Best Practices:}
\begin{itemize}
\item Reviewers insist on full and open access publication of source code, machine-readable training data, and artifacts such as trained models from publicly funded research.
\item Reviewers insist on task-appropriate performance indicators and fully-documented research protocol.
\item Identify and quantify bias / variance issues in datasets
\item Assess dataset and source bias through round-robin type studies to establish reproducible results.
\end{itemize}

\section{Opportunities in Workforce Development}
\label{sec-5}

There is an urgent need for workforce development to ensure that AI
techniques are introduced into the materials science workflow with the
appropriate level of scientific rigor. Briefly, there are opportunities
in:

\begin{enumerate}
\item Educating the next generation workforce to be conversant in AI techniques and their application to materials science.
\item Expanding skills within the current workforce, enabling them to  effectively mentor the next-generation workforce.
\end{enumerate}

Here consortia will play an important role by developing open source
educational materials, hosting bootcamps, and introducing workshop
tracks at professional meetings, creating learning opportunities and
materials that can be disseminated up and down the educational tiers.
Likewise the definition, publication, and demonstration of best
practices in AI will go a long way to increasing awareness and trust
within the community.

\subsection{Educating the next-generation workforce}
\label{sec-5-1}

A major limitation in the adoption of ML techniques by the materials
science community is a lack of an AI-savvy workforce. Traditional
materials science education contains few required courses in statistical
methods and programming. At the undergraduate level, there are few
treatments in the application of AI to materials science available for
developing course modules. One critical need is the development of open
data/code repositories that provide “plug and play” modules to augment
the current materials science undergraduate curriculum. These open
educational resources, in addition to formalized courses providing a
more rigorous introduction to research computing and statistical
methods, are needed to create a BS-level workforce capable of
implementing ML.

\subsection{Expanding skills within the current workforce}
\label{sec-5-2}

At the graduate and post-graduate level, there is an urgent need for
providing salient feedback on the relevance of models and their outputs.
Mid to late career materials scientists might feel unprepared to mentor
researchers applying ML techniques to their research. This can lead to
naively trusting (or dismissing out-of-hand) results from ML workflows,
or feeling unequipped to practice informed skepticism.

There is great need for a new professional track in the materials field,
since federally-funded data-intensive centers and facilities will start
building both physical and cyber data infrastructure. At present, the
availability of data technicians is minimal. Establishing a few training
pilots across the country amongst UG and CC will provide the needed
workforce to accomplish this task. If the MGI/AI/ OD/AM visions are to
be realized, there is an immediate urgency for workforce training
pilots.

\subsection{Paths Forward}
\label{sec-5-3}

\textbf{Consortia:}
\begin{itemize}
\item Develop open source educational materials (e.g.,  \href{https://datacarpentry.org/}{\url{https://datacarpentry.org/}} ) broadly targeting other opportunities throughout this document. Educational materials should leverage and reference known best practices.
\item Host bootcamps [e.g. NIST's MLMR], webinars, and hackathons framed  around AI usage in materials science
\item Introduce a “workshop track” at major materials conferences for  students and researchers to acquire and practice new skills.
\end{itemize}

\textbf{Best Practices:}
\begin{itemize}
\item Engage with stakeholders to define, publish, and demonstrate best  practices in the use of AI in Materials science and engineering.
\end{itemize}

\section{Summary and Outlook}
\label{sec-6}

In the previous sections, we provided a perspective on the application
of AI in materials science and engineering and outlined four overarching
opportunities for potential advancement within the community. We have
proposed five cross-cutting paths forward for each opportunity:

\textbf{Cross-disciplinary Collaboration} - Reinvigorated collaboration among the domains of materials science and engineering, computer science, and data science will advance state-of-the-art solutions for scientific AI and cyber-physical infrastructure while enabling trust  in AI.

\textbf{Autonomous Research Platforms} - The development and deployment of diverse autonomous research platforms will enable implementation and evaluation of new technology in scientific AI and cyber-physical  infrastructure by the rapid generation of high-quality experimental  materials data. Connecting these platforms will create compound  network effects that increase the leverage of any single experiment or calcualtion.

\textbf{Reference Data} - The creation of new reference and challenge datasets will enable the broader community to develop scientific AI and increase trust in AI, just as the classic MNIST handwritten digit database \cite{mnist} has enabled these outcomes in the broader STEM community.

\textbf{Consortia} - The creation of new consortia will engage stakeholders in industry, government and academia to provide economically sustainable frameworks for the deployment and operation of cyber-physical infrastructure and expanding the AI skills of the  current and future workforce, which will boost consortia member trust  in AI.

\textbf{Best Practices} - The creation of stakeholder-lead standards and best practices will enable trust in AI and foster a workforce that understands how to use AI effectively.

If the community makes coordinated efforts in these areas, we can
anticipate rapid acceleration of materials discovery and process
optimization, which will open new pathways for technological advancement
in sustainable development, transportation, water security, medicine,
and other technologies central to human welfare.

\printbibliography
% Emacs 25.3.1 (Org mode 8.2.10)
\end{document}